\RequirePackage{fix-cm}
\documentclass[smallextended]{svjour3}       
\smartqed  
\usepackage{graphicx}

\newif\iftikz
\tikztrue 

\usepackage{amsmath}
\iftikz
\usepackage{tikz}
\fi
\usepackage{graphicx}
\usepackage{hyperref}
\usepackage{epstopdf}

\usepackage{color}
\newcommand{\veps}{\ensuremath{\varepsilon}}
\newcommand{\ga}{\ensuremath{\alpha}}
\newcommand{\gb}{\ensuremath{\beta}}
\newcommand{\gc}{\ensuremath{\gamma}}
\newcommand{\gz}{\ensuremath{\zeta}}
\newcommand{\myi}{\ensuremath{\dot{\imath}}}
\newcommand{\ket}[1]{\ensuremath{\left|{#1}\right\rangle}}

\journalname{Quantum Information Processing}
\begin{document}

\title{Experimental pairwise entanglement estimation for an $N$-qubit system
}
\subtitle{A machine learning approach for programming quantum hardware}


\author{Nathan L. Thompson \and N.H. Nguyen \and  E. C. Behrman  \and
        James E. Steck
}


\institute{N.L. Thompson \and N.H. Nguyen \and E.C. Behrman \at
              Department of Mathematics, Statistics and Physics, Wichita State University, Wichita, KS 67260-0033 \\
              Tel.: +13169783953\\
              \email{behrman@math.wichita.edu}           
           \and
           J.E. Steck \at
              Department of Aerospace Engineering, Wichita State University, Wichita, KS 67260-0044
}

\date{Received: date / Accepted: date}

\maketitle

\begin{abstract}
Designing and implementing algorithms for medium and large scale quantum computers is not easy. In previous work we have suggested, and developed, the idea of using machine learning techniques to train a quantum system such that the desired process is ``learned,'' thus obviating the algorithm design difficulty. This works quite well for small systems. But the goal is macroscopic physical computation.  Here, we implement our learned pairwise entanglement witness on Microsoft's Q\#, one of the commercially available gate model quantum computer simulators; we perform statistical analysis to determine reliability and reproduceability; and we show that using the machine learning technique called ``bootstrapping'', we can infer the pattern for mesoscopic $N$ from simulation results for three-, four-, five-, six-, and seven-qubit systems. Our results suggest a fruitful pathway for general quantum computer algorithm design and computation.
\end{abstract}

\section{Introduction}
\label{intro}

For several decades now the prospect of macroscopic quantum computers, able to solve large classes of difficult problems, has been ``ten years away.'' We do have thousand-qubit size ``quantum annealing'' machines \cite{QA}, to solve optimization problems through adiabatic evolution to the ground state of a designed Hamiltonian, but programmable quantum computers remain small and their applicability limited. One major obstacle is the construction of algorithms that take advantage of the fundamental quantum nature of reality.  There are still only a very few. Most fall into one of three categories: those using a quantum Fourier transform, like Shor's algorithm \cite{shor}; those using amplitude amplification, like Grover's algorithm \cite{grover}; and those using quantum walks \cite{childs}. Speedup varies: Shor's, and some quantum walk algorithms, provide an exponential advantage over the best known classical algorithm in each case, but the speedup with Grover is only quadratic. We do not yet know whether there exists any quantum advantage for broad classes of problems \cite{bravyi,ronnow}, much less, what it will be in each case. Nor do we have a general process to factor an arbitrary $N$-qubit unitary efficiently to generate the quantum machine language necessary, in the case of the gate model; or to design a Hamiltonian whose ground state will be the answer to an optimization problem, in the case of quantum annealing.

For some time now our research group has been investigating the advantages of a marriage of machine learning and quantum computing to answer this need \cite{infsci,previous,2008}. The basic idea is that a quantum system can itself act as a neural network: The state of the system at the initial time is the ``input''; a measurement on the system at the final time is the ``output''. If we know enough about the computation desired to be able to construct a comprehensive set of input-output pairs from which the net can generalize, then, we can use techniques of machine learning to bypass the algorithm-construction problem.  Moreover, this approach is scalable \cite{multiqubit} using the machine learning technique called ``bootstrapping''\cite{bootstrap}, which uses knowledge of a smaller system to make systematic inferences about a larger one. In addition, our method promises to be generally robust to both noise and to decoherence \cite{2 qubit noise,robust}. Machine learning may also be helpful in the factorization problem \cite{related}, and in the Hamiltonian design problem \cite{LQA}.

Entanglement estimation is a good example of an intrinsically quantum calculation for which we have no general algorithm. Indeed, it has been shown that the quantum separability problem (determination of entanglement) is NP-hard \cite{gurvitz}. In previous work we succeeded in mapping a function of a measurement at the final time to a witness of the entanglement of a two-qubit system in its initial state \cite{2008}. The ``output'' (result of the measurement of the witness at the final time) will change depending on the time evolution of the system, which is of course controlled by the Hamiltonian: by the tunneling amplitudes $\{K\}$, the qubit biases $\{ \veps\}$, and the qubit-qubit coupling $\gz$. Thus we can consider these functions \{$K_A$, $K_B$, $\veps_A$, $\veps_B$, $\gz$\} to be the ``weights'' to be trained. We then use a quantum version \cite{2008} of backpropagation \cite{werbos} to find optimal functions such that our desired mapping is achieved.  Full details are provided in \cite{previous,2008}. From a training set of only four pure states, our quantum neural network successfully generalized the witness to large classes of states, mixed as well as pure \cite{multiqubit}. Qualitatively, what we are doing is using machine learning techniques to find a ``best'' hyperplane to divide separable states from entangled ones, in the Hilbert space.

Now, this method finds a time dependent Hamiltonian that solves the given problem, a procedure more reminiscent of a quantum annealing approach \cite{QA} than the gate approach. But of course the unitary operator of time development can be represented as a product of simple gates; indeed, it is a theorem \cite{neilsen and chuang} that any quantum computation can be performed as a succession of simple operators belonging to any universal set. Thus, a universal quantum computer need only be able to execute each of the members of that set \cite{lloyd}. There is now a large number of quantum simulators available online \cite{larose}, including Microsoft Quantum Development Kit \cite{Qsharp} and IBM's Quantum Experience \cite{IBMQE}, which implement a universal set of quantum operators (gates) plus many more that are useful in encoding quantum computations, such as the Pauli spin matrices, the Hadamard gate, the CNOT gate, and others.   The difficulty arises in determining exactly how to represent a particular calculation: first, in terms of finding the unitary for that problem; and second, in terms of these gates so that algorithms may be eventually implemented on real quantum hardware.

Once we do have a unitary, there are several approaches \cite{qcompiler,GateSynthesis} available for decomposing an arbitrary  $2^N\times 2^N$ unitary matrix, representing a quantum computation on an $N$-qubit system, into ``simple'' gates: single qubit and the two-qubit CNOT operations implemented in the languages associated with one of the online systems. However, none of these techniques is straightforward, and often the result is a large sequence of gates to represent the desired unitary. This inherent difficulty is another reason machine learning techniques are enticing \cite{pcoles}: If we can determine methods where the machines themselves develop and refine the algorithms they are using, we circumvent part of this intrinsic challenge of quantum computing. In this paper, because we want to demonstrate the advantages of bootstrapping, we will solve the two-qubit problem by hand, then bootstrap to generalize.

\section{Two-qubit Quantum Neural Network }

We begin with a simple two-qubit system.
\subsection{Reverse Engineering of Entanglement Witness}

The system evolves in time according to the Hamiltonian
\begin{equation} \label{Hamiltonian}
H = K_A X_A + K_B X_B + \veps_A Z_A + \veps_B Z_B + \gz Z_A Z_B
\end{equation}
where $X$ and $Z$ are the Pauli operators corresponding to qubits $A$ and $B$, $K_A$ and $K_B$ are the tunneling amplitudes, $\veps_A$ and $\veps_B$ are the biases, and $\gz$ is the qubit-qubit coupling.
The state of the system as a function of time can then be written, for a pure state, as
\begin{equation}\label{pureschr}
\ket{\psi(t)} = \exp\left({\dfrac{-\myi H t}{\hbar}}\right) \ket{\psi(0)}.
\end{equation}
It is convenient to consider the Hamiltonian $H$ as a sum of single qubit and two-qubit operations
\begin{equation} \label{Hbreakdown}
H = \displaystyle\underbrace{K_A X_A + \veps_A Z_A}_{H_A} + \underbrace{K_B X_B  + \veps_B Z_B}_{H_B} + \underbrace{\gz Z_A Z_B}_{H_{AB}}.
\end{equation}
We now consider the evolution to be broken into several ``time chunks''  where the parameters \{$K_A$, $K_B$, $\veps_A$, $\veps_B$, $\gz$\} are held constant on each interval. For most of the paper, we will use four time chunks or intervals. We can approximate the operator as the product of several operators as follows:
\begin{align} \label{expchunk}
\exp\left({\dfrac{-\myi H t}{\hbar}}\right) &= \prod_{k = 0}^3 \exp\left({\dfrac{-\myi H_k t}{4\hbar}}\right),
\end{align}
where on each time chunk $k$ the operator is approximated as
\begin{align} \label{expop}
\exp\left({\dfrac{-\myi H_k t}{4\hbar}}\right) &= \exp\left({\dfrac{-\myi (H_A + H_B + H_{AB})_k t}{4\hbar}}\right) \\
\label{expop2}
 &\approx \exp\left({\dfrac{-\myi H_{A,k} t}{4\hbar}}\right)\exp\left({\dfrac{-\myi H_{B,k} t}{4\hbar}}\right)\exp\left({\dfrac{-\myi H_{AB,k} t}{4\hbar}}\right).
\end{align}
Note that the last equation is only approximate, because while $H_A$ and $H_B$ commute, neither commutes with $H_{AB}$. During each interval or time chunk the functions \{$K_A$, $K_B$, $\veps_A$, $\veps_B$,  $\gz$\} are constant, so, we may, for a given time interval $\Delta t = t/4$, rewrite the operator given by (\ref{expop2}) as a product of physically realizable quantum gates, such as those implemented in the Q\# or Qiskit languages.

We start with the single qubit part of the operator for a single time interval, $\exp(-\myi H_A\Delta t/\hbar)$, and use the well-known identity for the exponential of Pauli matrices
\begin{equation} \label{Pauliexp}
e^{-\myi \ga (\bf{\hat{n}}\cdot\vec{\sigma})} = I\cos(\ga) - \myi(\bf{\bf{\hat{n}}}\cdot\vec{\sigma})\sin(\ga)
\end{equation}
where $I$ is the $2\times 2$ identity matrix, $\ga$ is an angle of rotation about axis $\bf{\hat{n}}$ (a unit vector) on the Bloch sphere, and $\vec{\sigma}$ is a vector of the Pauli matrices $\{X,Y,Z\}$. Looking at the definition of $H_A$ (or $H_B$) in equation (\ref{Hbreakdown}), we see that it is easy to express the exponent in the form of representation (\ref{Pauliexp}):
\begin{align}
\dfrac{\Delta t}{\hbar} H_A &= \dfrac{\Delta t}{\hbar}\left(K_A X_A + 0\, Y_A + \veps_A Z_A\right) \nonumber \\
        \label{HAPauli}     &= \displaystyle\underbrace{\dfrac{\Delta t}{\hbar}\sqrt{{K_A}^2 + {\veps_A}^2}}_\ga \underbrace{\left(\dfrac{K_A}{\sqrt{{K_A}^2 + {\veps_A}^2}} X_A + 0\, Y_A + \dfrac{\veps_A}{\sqrt{{K_A}^2 + {\veps_A}^2}} Z_A\right)}_{\bf{\hat{n}}\cdot\vec{\sigma}}.
\end{align}

Interpreting the operator as a rotation on the Bloch sphere, we have a formula for a rotation $\ga$ about an axis $\bf{\hat{n}}$ \cite{neilsen and chuang}
\begin{equation} \label{rotation}
R_{\bf{\hat{n}}} (\ga) = R_z(\gc)R_y(\gb)R_z(\ga)R_y(-\gb)R_z(-\gc),
\end{equation}
where the rotations $R_x(\theta)$, $R_y(\theta)$, and $R_z(\theta)$ are defined as
\begin{align*}
R_x(\theta) &= e^{-\myi\frac{\theta}{2}X} = \left[\begin{tabular}{cc}
                                                $\cos\frac{\theta}{2}$ & $-\myi\sin\frac{\theta}{2}$ \\
                                                $-\myi\sin\frac{\theta}{2}$ & $\cos\frac{\theta}{2}$
                                                \end{tabular} \right] \\
R_y(\theta) &= e^{-\myi\frac{\theta}{2}Y} = \left[\begin{tabular}{cc}
                                                $\cos\frac{\theta}{2}$ & $-\sin\frac{\theta}{2}$ \\
                                                $-\sin\frac{\theta}{2}$ & $\cos\frac{\theta}{2}$
                                                \end{tabular} \right] \\
R_z(\theta) &= e^{-\myi\frac{\theta}{2}Z} = \left[\begin{tabular}{cc}
                                                $e^{-\myi\frac{\theta}{2}}$ & 0 \\
                                                0 & $e^{\myi\frac{\theta}{2}}$
                                                \end{tabular} \right]
\end{align*}
The Q\# and Qiskit environments \cite{Qsharp,IBMQE} have access to a function which computes the rotation of a state about the $x$, $y$, or $z$ axis of the Bloch sphere by a specified angle, so this expression will suffice supposing that we can find the appropriate values for $\ga$, $\gb$, and $\gc$ in (\ref{rotation}). To do this, we use some analogues to this expression in terms of Pauli matrices and spherical coordinates:
\begin{align} \label{RotPauli}
R_{\bf{\hat{n}}} (\ga) &= I\cos\left(\frac{\ga}{2}\right) - \myi\left(\bf{\hat{n}}\cdot\vec{\sigma}\right)\sin\left(\frac{\ga}{2}\right) \nonumber \\
                  &= I\cos\left(\frac{\ga}{2}\right) - \myi\left(\sin\gb\cos\gc\, X + \sin\gb\sin\gc\, Y + \cos\gb\, Z\right)\sin\left(\frac{\ga}{2}\right).
\end{align}

Our expression (\ref{HAPauli}) matches (\ref{RotPauli}) perfectly, and now we need only solve the following system of three equations with three unknowns:
\begin{equation} \label{findnhat}
\sin\gb\cos\gc = \dfrac{K_A}{\sqrt{{K_A}^2 + {\veps_A}^2}}, \quad \sin\gb\sin\gc = 0, \quad \cos\gb  = \dfrac{\veps_A}{\sqrt{{K_A}^2 + {\veps_A}^2}}.
\end{equation}
We notice immediately that $\sin\gc = 0$ (since $\sin\gb$ cannot be zero due the first equation), and so $\gc = c\pi$ for some integer $c$. This forces $\cos\gc$ to be $\pm 1$. Last, $\gb = \sin^{-1}\left(\pm K_A/\sqrt{{K_A}^2 + {\veps_A}^2}\right)$. We see that the relative sizes of $K_A$ and $\veps_A$ are constrained by the sine and cosine relationship between
\begin{equation}
\sin\gb = \pm\dfrac{K_A}{\sqrt{{K_A}^2 + {\veps_A}^2}} \qquad \cos\gb  = \dfrac{\veps_A}{\sqrt{{K_A}^2 + {\veps_A}^2}}. \nonumber
\end{equation}
A change of indices gives us the operator $H_B$  in a similar way. The only remaining step is to write $H_{AB}$ of (\ref{Hbreakdown}) in a form using practical quantum gates.

The two-qubit part of the Hamiltonian is $H_{AB} = \gz Z_A Z_B$. The matrix form of this operator is generated by taking the Kronecker product, $Z_A Z_B = Z\otimes Z$.  After setting $w_0 = \gz\Delta t/\hbar$ and taking the exponential of the operator, we have
\begin{equation} \label{HABmatrix}
\exp\left(\dfrac{-\myi H_{AB} \Delta t}{\hbar}\right) = \left[\begin{tabular}{cccc}
    $e^{ -\myi w_0}$ & 0 & 0 & 0 \\
    0 & $e^{ \myi w_0}$ & 0 & 0 \\
    0 & 0 & $e^{ \myi w_0}$ & 0 \\
    0 & 0 & 0 & $e^{ -\myi w_0}$
\end{tabular} \right].
\end{equation}
Since this is a two-qubit operator, it is necessary to represent it using a two-qubit gate. The primary tool for this is the CNOT (controlled NOT) gate,
\begin{equation} \label{CNOT}
\text{CNOT} = \left[\begin{tabular}{cccc}
    1 & 0 & 0 & 0 \\
    0 & 1 & 0 & 0 \\
    0 & 0 & 0 & 1 \\
    0 & 0 & 1 & 0
\end{tabular} \right].
\end{equation}
In general the CNOT operator in addition to single-qubit phase gates forms a universal set with which to build an arbitrary ($N$-qubit) operator. For our purposes, we may represent matrix (\ref{HABmatrix}) using the following expression:

\begin{equation} \label{twoqubit}
\exp\left(\dfrac{-\myi H_{AB} \Delta t}{\hbar}\right) = \text{CNOT}
\left[\begin{tabular}{cccc}
    $e^{-\myi w_0}$ & 0 & 0 & 0 \\
    0 & $e^{ \myi w_0}$ & 0 & 0 \\
    0 & 0 & $e^{-\myi w_0}$ & 0 \\
    0 & 0 & 0 & $e^{ \myi w_0}$
\end{tabular} \right] \text{CNOT}.
\end{equation}
The interior matrix is $I \otimes R_z(2 w_0)$, which is a rotation on only the $B$ qubit. With the above decompositions for $H_A$, $H_B$, and $H_{AB}$, we can now express each time chunk of our quantum operator in terms of a quantum circuit

\iftikz
\vspace{0.1cm}
\noindent\scalebox{0.8}{\providecommand{\ket}[1]{\left|#1\right\rangle}
\begin{tikzpicture}[scale=1.500000,x=1pt,y=1pt]
\filldraw[color=white] (0.000000, -7.500000) rectangle (249.000000, 22.500000);
\draw[color=black] (0.000000,15.000000) -- (249.000000,15.000000);
\draw[color=black] (0.000000,15.000000) node[left] {$\ket{A}$};
\draw[color=black] (0.000000,0.000000) -- (249.000000,0.000000);
\draw[color=black] (0.000000,0.000000) node[left] {$\ket{B}$};
\draw (9.000000,15.000000) -- (9.000000,0.000000);
\begin{scope}
\draw[fill=white] (9.000000, 0.000000) circle(3.000000pt);
\clip (9.000000, 0.000000) circle(3.000000pt);
\draw (6.000000, 0.000000) -- (12.000000, 0.000000);
\draw (9.000000, -3.000000) -- (9.000000, 3.000000);
\end{scope}
\filldraw (9.000000, 15.000000) circle(1.500000pt);
\begin{scope}
\draw[fill=white] (39.000000, 0.000000) +(-45.000000:21.213203pt and 8.485281pt) -- +(45.000000:21.213203pt and 8.485281pt) -- +(135.000000:21.213203pt and 8.485281pt) -- +(225.000000:21.213203pt and 8.485281pt) -- cycle;
\clip (39.000000, 0.000000) +(-45.000000:21.213203pt and 8.485281pt) -- +(45.000000:21.213203pt and 8.485281pt) -- +(135.000000:21.213203pt and 8.485281pt) -- +(225.000000:21.213203pt and 8.485281pt) -- cycle;
\draw (39.000000, 0.000000) node {$R_z(2w_{5k})$};
\end{scope}
\draw (69.000000,15.000000) -- (69.000000,0.000000);
\begin{scope}
\draw[fill=white] (69.000000, 0.000000) circle(3.000000pt);
\clip (69.000000, 0.000000) circle(3.000000pt);
\draw (66.000000, 0.000000) -- (72.000000, 0.000000);
\draw (69.000000, -3.000000) -- (69.000000, 3.000000);
\end{scope}
\filldraw (69.000000, 15.000000) circle(1.500000pt);
\begin{scope}
\draw[fill=white] (106.500000, 15.000000) +(-45.000000:31.819805pt and 8.485281pt) -- +(45.000000:31.819805pt and 8.485281pt) -- +(135.000000:31.819805pt and 8.485281pt) -- +(225.000000:31.819805pt and 8.485281pt) -- cycle;
\clip (106.500000, 15.000000) +(-45.000000:31.819805pt and 8.485281pt) -- +(45.000000:31.819805pt and 8.485281pt) -- +(135.000000:31.819805pt and 8.485281pt) -- +(225.000000:31.819805pt and 8.485281pt) -- cycle;
\draw (106.500000, 15.000000) node {$R_y(-w_{5k+1})$};
\end{scope}
\begin{scope}
\draw[fill=white] (106.500000, 0.000000) +(-45.000000:31.819805pt and 8.485281pt) -- +(45.000000:31.819805pt and 8.485281pt) -- +(135.000000:31.819805pt and 8.485281pt) -- +(225.000000:31.819805pt and 8.485281pt) -- cycle;
\clip (106.500000, 0.000000) +(-45.000000:31.819805pt and 8.485281pt) -- +(45.000000:31.819805pt and 8.485281pt) -- +(135.000000:31.819805pt and 8.485281pt) -- +(225.000000:31.819805pt and 8.485281pt) -- cycle;
\draw (106.500000, 0.000000) node {$R_y(-w_{5k+2})$};
\end{scope}
\begin{scope}
\draw[fill=white] (163.500000, 15.000000) +(-45.000000:31.819805pt and 8.485281pt) -- +(45.000000:31.819805pt and 8.485281pt) -- +(135.000000:31.819805pt and 8.485281pt) -- +(225.000000:31.819805pt and 8.485281pt) -- cycle;
\clip (163.500000, 15.000000) +(-45.000000:31.819805pt and 8.485281pt) -- +(45.000000:31.819805pt and 8.485281pt) -- +(135.000000:31.819805pt and 8.485281pt) -- +(225.000000:31.819805pt and 8.485281pt) -- cycle;
\draw (163.500000, 15.000000) node {$R_z(w_{5k+3})$};
\end{scope}
\begin{scope}
\draw[fill=white] (163.500000, 0.000000) +(-45.000000:31.819805pt and 8.485281pt) -- +(45.000000:31.819805pt and 8.485281pt) -- +(135.000000:31.819805pt and 8.485281pt) -- +(225.000000:31.819805pt and 8.485281pt) -- cycle;
\clip (163.500000, 0.000000) +(-45.000000:31.819805pt and 8.485281pt) -- +(45.000000:31.819805pt and 8.485281pt) -- +(135.000000:31.819805pt and 8.485281pt) -- +(225.000000:31.819805pt and 8.485281pt) -- cycle;
\draw (163.500000, 0.000000) node {$R_z(w_{5k+4})$};
\end{scope}
\begin{scope}
\draw[fill=white] (220.500000, 15.000000) +(-45.000000:31.819805pt and 8.485281pt) -- +(45.000000:31.819805pt and 8.485281pt) -- +(135.000000:31.819805pt and 8.485281pt) -- +(225.000000:31.819805pt and 8.485281pt) -- cycle;
\clip (220.500000, 15.000000) +(-45.000000:31.819805pt and 8.485281pt) -- +(45.000000:31.819805pt and 8.485281pt) -- +(135.000000:31.819805pt and 8.485281pt) -- +(225.000000:31.819805pt and 8.485281pt) -- cycle;
\draw (220.500000, 15.000000) node {$R_y(w_{5k+1})$};
\end{scope}
\begin{scope}
\draw[fill=white] (220.500000, 0.000000) +(-45.000000:31.819805pt and 8.485281pt) -- +(45.000000:31.819805pt and 8.485281pt) -- +(135.000000:31.819805pt and 8.485281pt) -- +(225.000000:31.819805pt and 8.485281pt) -- cycle;
\clip (220.500000, 0.000000) +(-45.000000:31.819805pt and 8.485281pt) -- +(45.000000:31.819805pt and 8.485281pt) -- +(135.000000:31.819805pt and 8.485281pt) -- +(225.000000:31.819805pt and 8.485281pt) -- cycle;
\draw (220.500000, 0.000000) node {$R_y(w_{5k+2})$};
\end{scope}
\end{tikzpicture}
}
\vspace{0.1cm}
\fi

\noindent where we have relabeled variables as $w_{5k} = \dfrac{\gz_k\Delta t}{\hbar}$, \\
\noindent $w_{5k+1} = \sin^{-1}\left(\dfrac{K_{A,k}}{\sqrt{K_{A,k}^2+\veps_{A,k}^2}}\right)$,\ $w_{5k+2} = \sin^{-1}\left(\dfrac{K_{B,k}}{\sqrt{K_{B,k}^2+\veps_{B,k}^2}}\right)$, \\
$w_{5k+3} = \dfrac{\Delta t}{\hbar}\sqrt{K_{A,k}^2+\veps_{A,k}^2}$, and $w_{5k+4} = \dfrac{\Delta t}{\hbar}\sqrt{K_{B,k}^2+\veps_{B,k}^2}$.

\vspace{0.3cm}
\noindent Collected formulaically, the gate decomposition of operator (\ref{expchunk}) is
\begin{align} \label{gatechunk}
\exp\left({\dfrac{-\myi H t}{\hbar}}\right) &= \prod_{k=0}^3 U_{A,k}\, U_{B,k}\, U_{AB,k}
\end{align}
where
\begin{align}
\label{UA}
U_{A,k} &= \left[ R_y(w_{5k+1}) R_z(w_{5k+3}) R_y(-w_{5k+1}) \right] \otimes I  \\
\label{UB}
U_{B,k} &= I \otimes \left[ R_y(w_{5k+2}) R_z(w_{5k+4}) R_y(-w_{5k+2}) \right]  \\
\label{UAB}
U_{AB,k} &= \text{CNOT} \left[ I \otimes R_z (2w_{5k}) \right] \text{CNOT}.
\end{align}

\subsection{Numerical computation}

In our original work \cite{previous,2008,multiqubit} on the entanglement witness, we used piecewise constant functions for \{$K_A$, $K_B$, $\veps_A$, $\veps_B$, $\gz$\}; in subsequent work \cite{continuous} we used continuum functions, for which we found training was much more rapid and complete. Because current technology does not allow for continuous-time control of gate functions, we return to our original piecewise formulation; however, we have retrained using our more recent codes to improve our earlier results.

Physically, we imagine that the system would be allowed to evolve for a specified time under a Hamiltonian whose parameter functions we could control. At the end of that time we perform a measurement whose average value would represent the entanglement witness. The training of the net is a process whereby we find an optimal mapping of the desired physical property (here, the entanglement) to that chosen measurement. We chose, as that measurement, the (square of the) qubit-qubit correlation function at that final time, $\langle Z_{A} (t_{f}) Z_{B}(t_{f}) \rangle^{2}$.

To perform the retraining, we used our (newer) continuum codes, but averaged each parameter function over each time interval, then used that averaged  function as the piecewise constant parameter for time evolution, to calculate the expectation value of the final time correlation function, and, therefore, the error. Training data are shown in Table \ref{enttraintable}. (See \cite{2008} for full details.)  The ``Desired'' column is the goal of the training for the final time correlation function, showing that we seek a value of one for a fully entangled state and zero for a product state. Because we are trying to optimize our entanglement witness, we find a value intermediate between zero and one for the target value for the partially entangled state state $|$P$\rangle$; this (optimized) value is 0.443.  The column labelled ``Trained'' shows the asymptotic value for that final time correlation function after the training of the network. Training was of course less efficient than with the greater flexibility offered by the continuous-time functions; nonetheless, RMS error for the training set was only 0.05\% after 200 epochs. The piecewise constant values found for the parameter functions are shown in Table \ref{paramtable}.

\vspace*{4pt}   
\begin{table}[htbp]
\caption{QNN entanglement witness trained for 200 epochs using piecewise constant parameter functions, and compared with calculated results using first chunked time propagators, then with the sequence of gates, and finally on the Q\# simulator \cite{Qsharp}. The training set of four \cite{2008} includes one completely entangled state $|\text{Bell}\rangle = \frac{1}{\sqrt{2}}[|00\rangle + |11\rangle]$, one unentangled state $|\text{Flat}\rangle = \frac{1}{2}[|00\rangle+|01\rangle+|10\rangle+|11\rangle]$, one classically correlated but unentangled state $|\text{C}\rangle = \frac{1}{\sqrt{5}}[2|00\rangle+|01\rangle]$, and one partially entangled state $|\text{P}\rangle = \frac{1}{\sqrt{3}}[|01\rangle+|10\rangle+|11\rangle]$. Errors for each method are shown in the final line.}

\begin{tabular}{l l l l l l}\\
\hline
Input state & Desired & Trained & Chunked & Gates & Q\# \\
\hline
$|\text{Bell}\rangle$ & 1.0 & 0.999 & 0.999 & 0.999 & 0.999 \\
$|\text{Flat}\rangle$ & 0.0 & $7.99 \times 10^{-5}$  &  $5.99 \times 10^{-7}$ & $5.99 \times 10^{-7}$ &  $6.14 \times 10^{-5}$ \\
$|\text{C}\rangle$ & 0.0 & $1.08 \times 10^{-4}$  &  $1.87 \times 10^{-5}$  & $1.87 \times 10^{-5}$ & $8.17 \times 10^{-5} $ \\
$|\text{P}\rangle$ & 0.443  & 0.440  & 0.446 & 0.446 & 0.446      \\
\hline
Total RMS error   &   { }  & $5.0 \times 10^{-4}$ & $1.4 \times 10^{-3}$ & $1.4 \times 10^{-3}$  & $1.7 \times 10^{-3}$ \\
\hline \\
\end{tabular} 
\label{enttraintable}
\end{table}

\vspace*{4pt}   
\begin{table}[htbp]
\caption{Trained parameter functions for the entanglement witness for the two-qubit system, in MHz.  Total time of evolution for the two time propagation methods was 1.58 ns.  }
\begin{tabular}{l c c c c}\\
\hline
Parameter &   Interval 1 & Interval 2 & Interval 3 & Interval 4   \\
\hline
$K_{A} = K_{B} $  &  2.49  &   2.47 & 2.48 & 2.49  \\
$\zeta$        &  0.0382   &  0.128 & 0.117 & 0.0382  \\
$\veps_{A}=\veps_{B}$   &   0.0930  & 0.116 & 0.0954 & 0.0833 \\
\hline \hline \\
\end{tabular}
\label{paramtable}
\end{table}

We now use these trained values for the piecewise constant parameter functions in the equations derived for the sequence of operators in the previous section. Note that there are two separate sources for the error: the approximation in Equation (\ref{expop2}), which assumes that the matrices commute; and the approximation of the substitution of the products of the gate operators for the time-propagation operator. We can separate these two sources by calculating the density matrix for the final time, using ``chunked time.'' That is, instead of calculating the time propagation correctly as in the QNN training, we separate the Hamiltonian into $H_A$, $H_B$, and $H_{AB}$ for each of the four time intervals. This assumes that the pieces commute, which of course is an approximation. The column in Table \ref{enttraintable} labelled ``Chunked,'' shows the calculation of the entanglement witness using this approximation. The ``Gates'' column repeats these calculations using the matrix decomposition outlined in Section 2.1. The final column, labelled ``Q\#'', shows the entanglement witness values of the sequence of gates as measured on Microsoft's quantum simulator \cite{Qsharp}. (Calculations performed using IBM's Quantum Experience simulator \cite{IBMQE} produced almost identical results \cite{AIAA}.) Note that the calculated numbers for the entanglement witness in the two last columns are extremely close, as are, of course, the RMS errors for each method. The Frobenius norm of the difference between the density matrix as trained by the QNN technique and the (non-commuting) chunked time propagation matrices is in each case 1 to 2\%; while the norm of the difference between the density matrix calculated by the chunked time propagator and by the sequence of applied gates in each case is on the order of $10^{-15}$, i.e., within round-off error. Clearly all of the error comes from the non-commutation. This validates our replacement of the time evolution operator by the product of gates.

\section{Statistical Evaluation of Entanglement Witness in Q\#}

With the entanglement witness properly reverse engineered to run on the hardware simulators, we now need to understand how to utilize it in applied situations. Both the Q\# and Qiskit systems implement measurements of the qubit along a standard axis $x, y,$ or $z$ in the Bloch sphere. Each individual measurement only returns an eigenvalue of $\pm 1$. To generate a useable expected value, several thousand measurements must be done to average these eigenvalues to get a valid approximation of the expectation value $\langle Z_{A} (t_{f}) Z_{B}(t_{f}) \rangle^{2}$. Using the Q\# built-in simulator, we did 100 iterations at several different ``shot counts'' (number of individual experiments and measurements) to gauge how many times a particular experiment must be run to generate a high confidence value for the entanglement witness. Our code is available at \cite{github}.

\begin{figure}[h]
    \begin{center}
        \includegraphics[height=4.5cm]{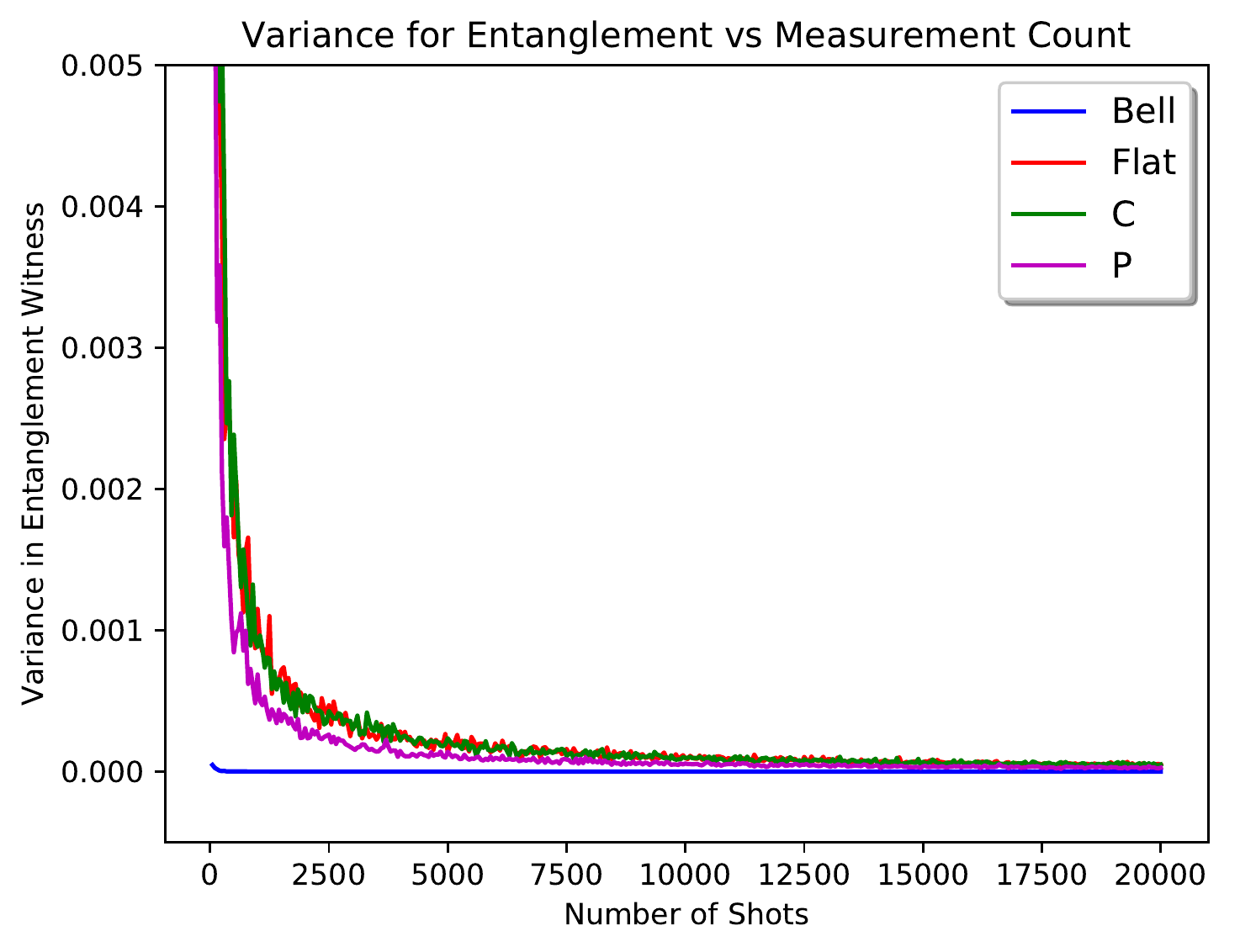}
    \end{center}
    \caption{Variance in entanglement witness for 100 iterations of each state measured at shot counts ranging from 50 to 20,000 in 50 shot increments. As the shot count increases, we see that the measurement variance quickly goes to zero.}
        \label{VariancePlot}
\end{figure}

\begin{figure}[htbp]
    \begin{minipage}{.5\textwidth}
        \includegraphics[height = 4.5cm]{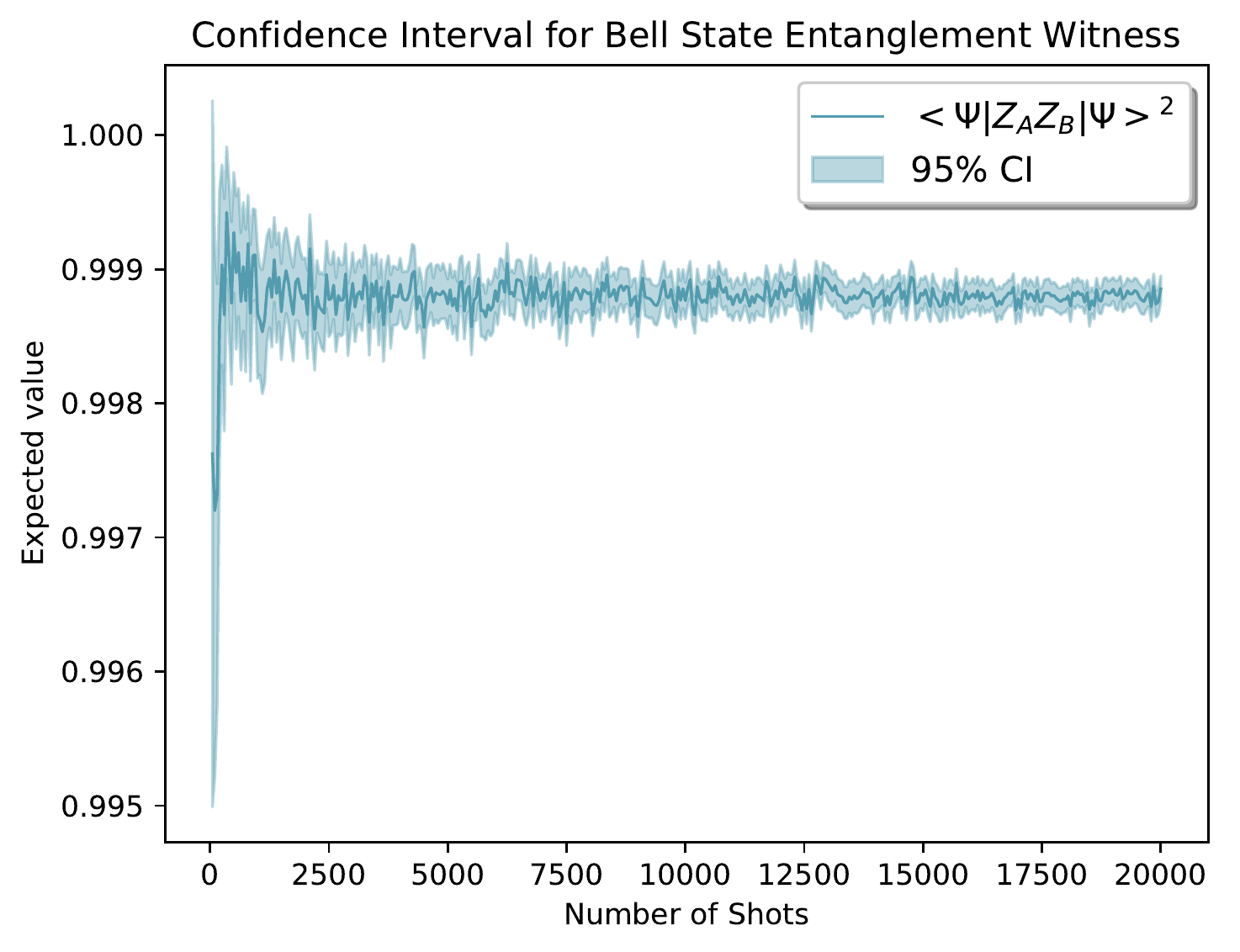}
    \end{minipage}
    \begin{minipage}{.5\textwidth}
        \includegraphics[height = 4.5cm]{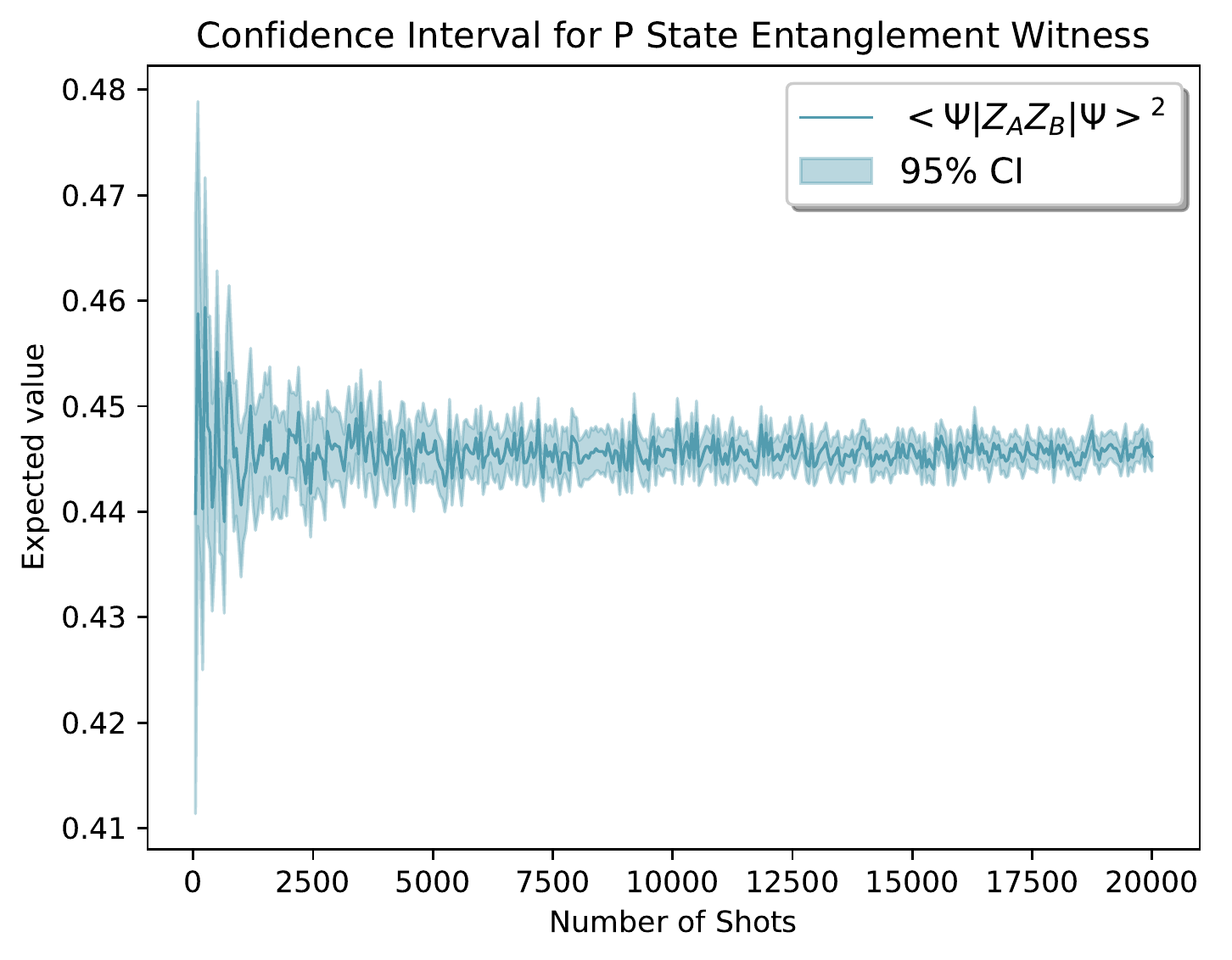}
    \end{minipage}
    \caption{Q\# entanglement witness values for the $|$Bell$\rangle$ and $|$P$\rangle$ states with a 95\% confidence interval as a function of the shot count. The confidence interval (CI) width reaches its minimum of $\sim$0.0015 after approximately 15,000 shots.}
        \label{BellP_conf}
\end{figure}

Figure \ref{VariancePlot} shows the variance of the expectation value $\langle Z_{A} (t_{f}) Z_{B}(t_{f}) \rangle^{2}$ as a function of numbers of shot counts. We can see plainly that the law of large numbers is in effect for determining the entanglement witness. High confidence values for the witness are achieved near 15,000 iterations of the experiment. This is easier to see in Figure \ref{BellP_conf}, which shows a 95\% confidence interval surrounding the computed square of the qubit-qubit correlation for the witness on the $|$Bell$\rangle$ and $|$P$\rangle$ states, where the width of the interval shrinks to 0.0015. Results for the $|$Flat$\rangle$ and $|$C$\rangle$ states are similar.

\section{Bootstrapping}

We have constructed a sequence of hardware gates that mimics our trained two-qubit entanglement witness quite well. While this is interesting it is perhaps of somewhat limited use, as it pertains only to a two-qubit system. We now extend our results to an $N$-qubit system.

\subsection{Searching for an Asymptotic Limit}

\begin{figure}[h]
    \begin{minipage}{.5\textwidth}
        \includegraphics[height = 4.5cm]{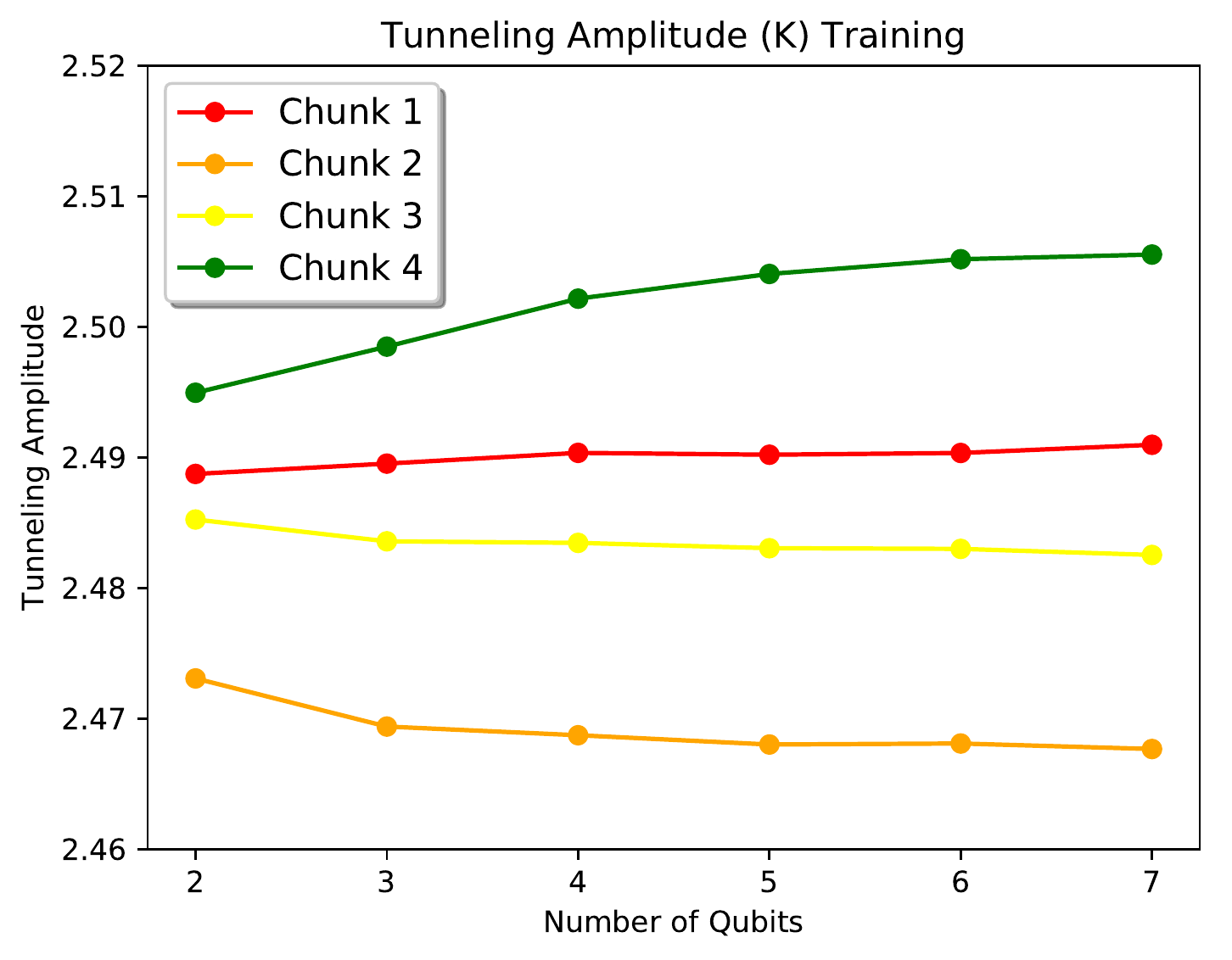}
    \end{minipage}
    \begin{minipage}{.5\textwidth}
        \includegraphics[height = 4.5cm]{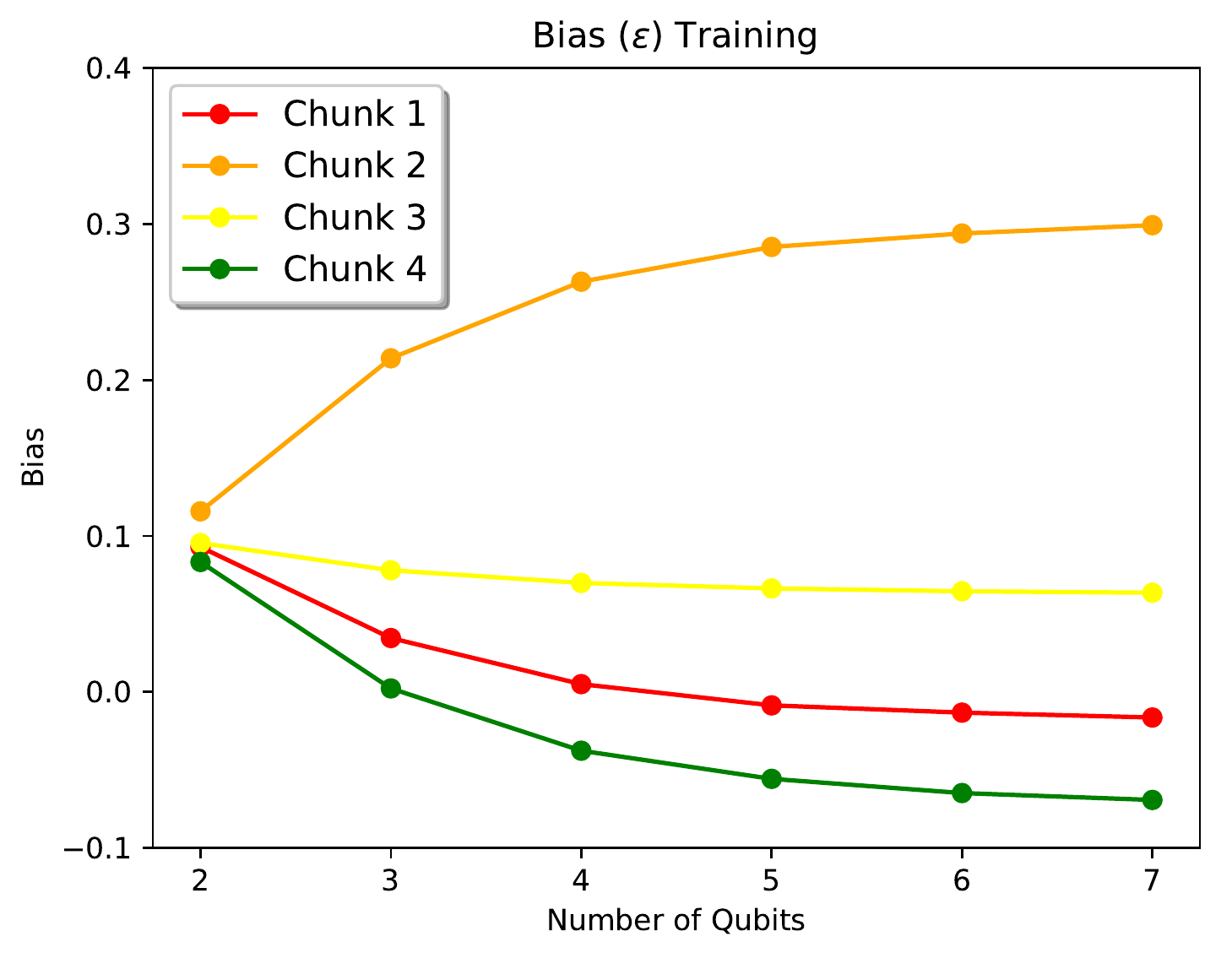}
    \end{minipage}
        \caption{Trained values for the tunneling amplitude $K$ and for the bias $\veps$, for each time chunk, as the number of qubits in the system is increased. Both demonstrate clear asymptotic behavior.}
        \label{asymptrain1}
\end{figure}

\begin{figure}[h]
    \begin{center}
        \includegraphics[height = 4.5cm]{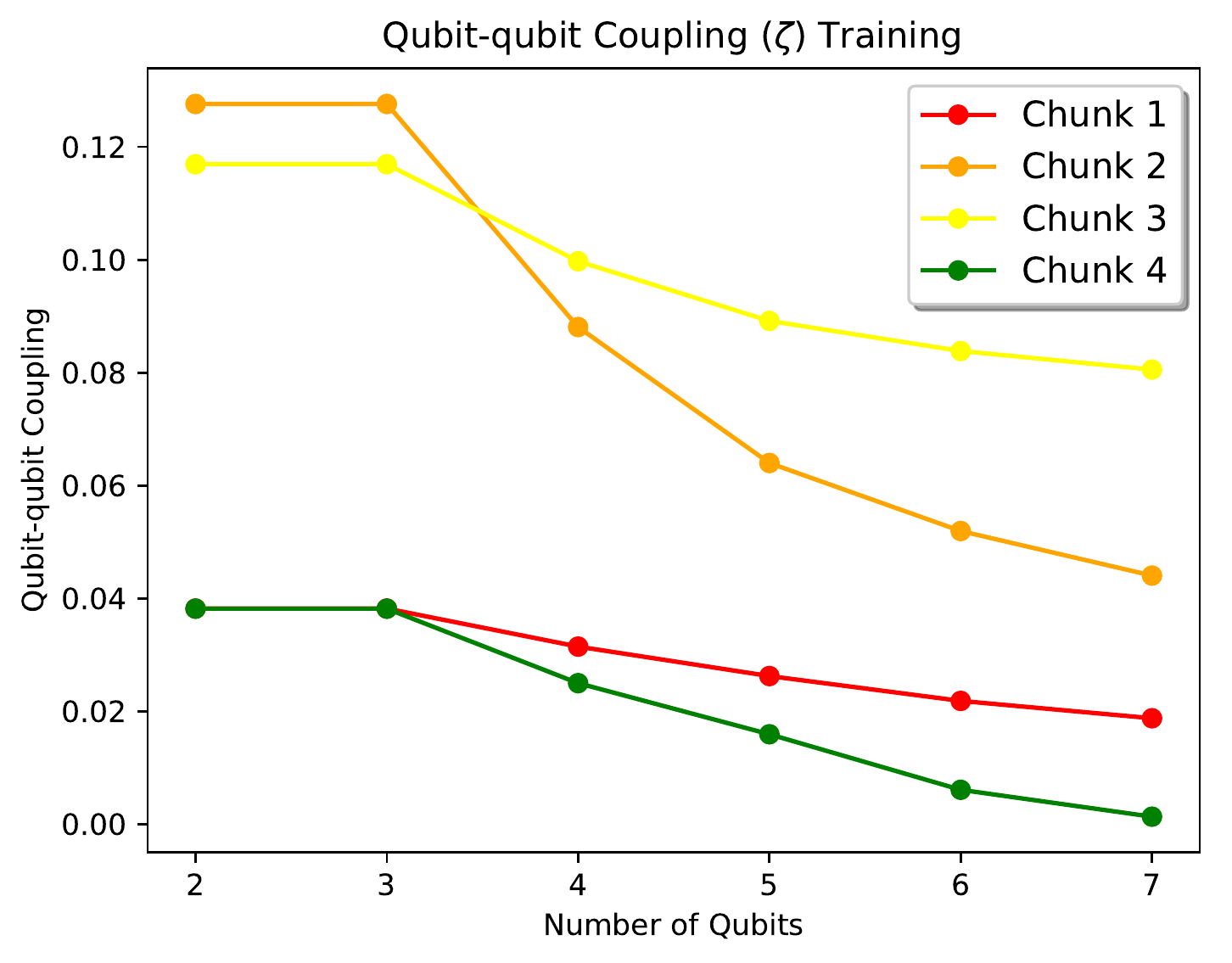}
    \end{center}
    \caption{Trained values for the qubit-qubit coupling $\zeta$, for each time chunk, as the number of qubits in the system is increased. The values show a clear trend, but do not become asymptotic as quickly as with the other parameters.}
    \label{asymptrain2}
\end{figure}

The technique of ``bootstrapping'' \cite{bootstrap} involves using knowledge of a smaller system to make systematic inferences about a larger, or, to use a partial knowledge of a system to infer a greater. So, as an initial guess for the correct parameter functions for the three-qubit system, we take the parameter functions \{$K_A = K_B$, $\veps_A = \veps_B$, $\gz$\} for the two-qubit system. We then train the three-qubit system from that point to minimize the error.
Once the three-qubit trained functions are found, we start from those to train the four-qubit system, and so on. The benefit is that while there are large changes in the tunneling, bias, and coupling parameters as the system size increases initially, those percentage changes diminish as the system size $N$ increases, due to the increased connectivity. Hence, training five-, six-, and seven-qubit systems require fewer and fewer additional epochs. Because of the symmetry of the problem, all the parameter functions can be taken to be the same (that is, $K_{A}(t)=K_{B}(t)=K_{C}(t)$, $\veps_{A}(t)=\veps_{B}(t)=\veps_{C}(t)$, and so on); imposing this as a constraint also reduces the training time. We now look for an asymptotic limit as $N$ increases. Figures \ref{asymptrain1} and \ref{asymptrain2} show the results of the training.

\vspace*{4pt}   

All parameters show an asymptotic trend, with the tunneling amplitudes $K$ and biases $\veps$ showing swift convergence to a limiting value. The qubit-qubit coupling $\zeta$ also has a trend emerging at the number of qubits increases, indicating that an $N$-qubit limit is likely. We infer that the parameters for the seven-qubit system are a reasonable approximation for the entanglement witness of an $N$-qubit system based on the limiting behavior observed in $K$ and $\veps$. This is important, because once quantum computers become only a very little larger we will no longer be able to simulate them on classical computers (the point of so-called ``quantum supremacy'' \cite{preskill}.) Table \ref{multitrain} contains the parameters for the fully symmetric seven-qubit system.

\begin{table}[h]
\caption{Trained parameter values at each time interval, for the pairwise entanglement witness for the seven-qubit system, in MHz. By symmetry, each of the tunneling functions $K$, each of the biases $\veps$, and each of the pairwise couplings $\zeta$ is the same. We take these values to be an approximation to the asymptotic limit of the parameters for an $N$-qubit quantum system. }
\begin{tabular}{l c c c c}\\
\hline
Parameter &   Interval 1 & Interval 2 & Interval 3 & Interval 4   \\
\hline
$K $  &  2.49   & 2.47  & 2.48  & 2.51  \\
$\zeta$        &  0.0188  & 0.0440 & 0.0805 & 0.00132 \\
$\veps$   &  -0.0164  & 0.299 & 0.0636 & -0.0693 \\
\hline \hline \\
\end{tabular}
\label{multitrain}
\end{table}

Training for these parameters was relatively efficient for an $N$-qubit system, taking only 100 additional epochs (passes through the whole training set) past the previously trained $(N-1)-$qubit system to train the pairwise entanglement witness. While the number of training pairs,  $4\binom{N}{2}$, does increase with the number of qubits,  the increased connectivity meant that the system needed less additional training each time.  The total RMS value for the training of the two-qubit parameters is $6.0\times 10^{-4}$, and only increased slightly as qubits were added, with six-qubits having an RMS of $1.6\times 10^{-3}$ (at 60 training pairs) and seven-qubits $1.8\times 10^{-3}$  (84 training pairs). Mesoscopic systems will still require some training to decrease initial errors, but this amount should be very small or negligible since we already see the parameters nearing asymptotic values, and we anticipate that this (small) additional training can be done on-line and need not be simulated.

\subsection{Comparing the Discrete and Continuum Cases}

Having established the scalability of our results in terms of growing system size, we now show how the results for the chunked system compare to our more sophisticated model for the entanglement witness studied in \cite{continuous}. In that work, the tunneling, bias, and coupling parameters were all modeled using continuous functions of time. Allowing continuous parameters added a great deal more flexibility and assisted training immensely. This approach to quantum machine learning resulted in smaller errors and faster training than the piecewise constant ``chunked'' model. However, with the current gate based model of quantum computing, we have no expectation of being able to implement or train a continuum parameter solution on developing or proposed hardware. Therefore, we examine the relationship between entanglement witnesses built and trained using the chunked and continuous versions of the $K$, $\veps$, and $\zeta$ parameters.

\begin{figure}[!h]
    \begin{minipage}{.5\textwidth}
        \includegraphics[height = 4.5cm]{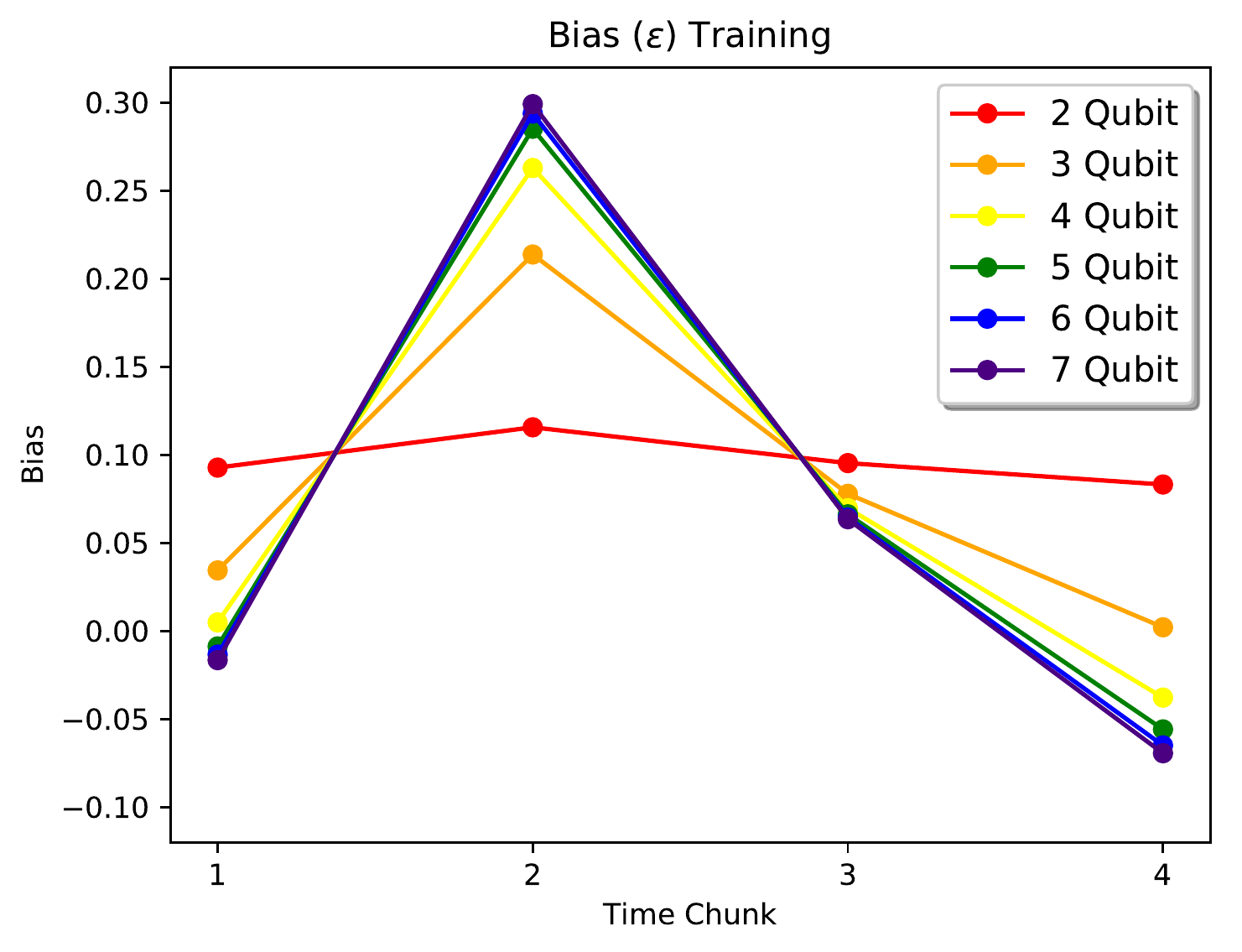}
    \end{minipage}
    \begin{minipage}{.5\textwidth}
        \includegraphics[height = 4.5cm]{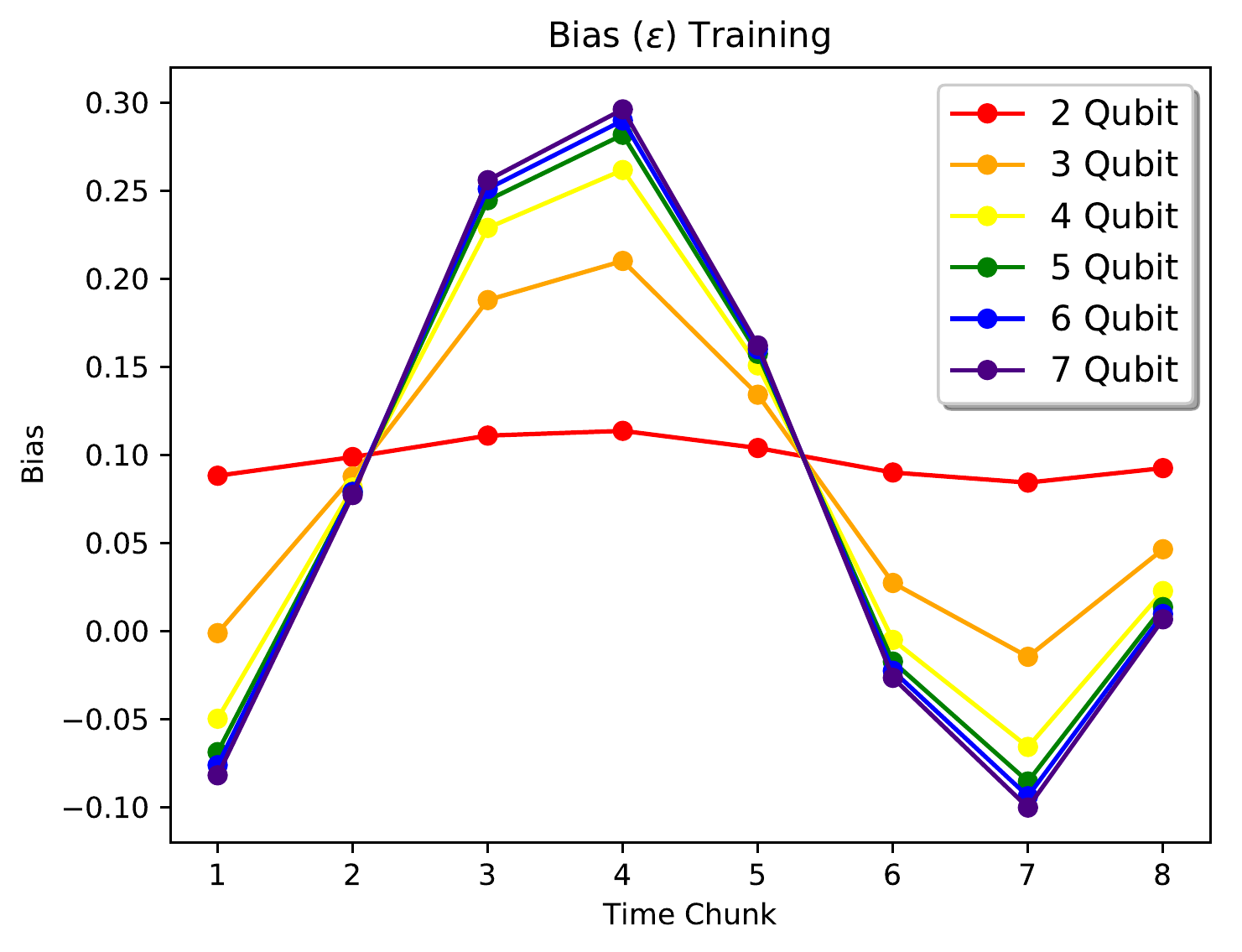}
    \end{minipage}
    \caption{Trained bias, for 4 and 8 time chunks, as functions of time, for systems of increasing numbers of qubits $N$.}
    \label{Eps_Chunk}
\end{figure}
\begin{figure}
    \begin{center}
        \includegraphics[height = 4.5cm]{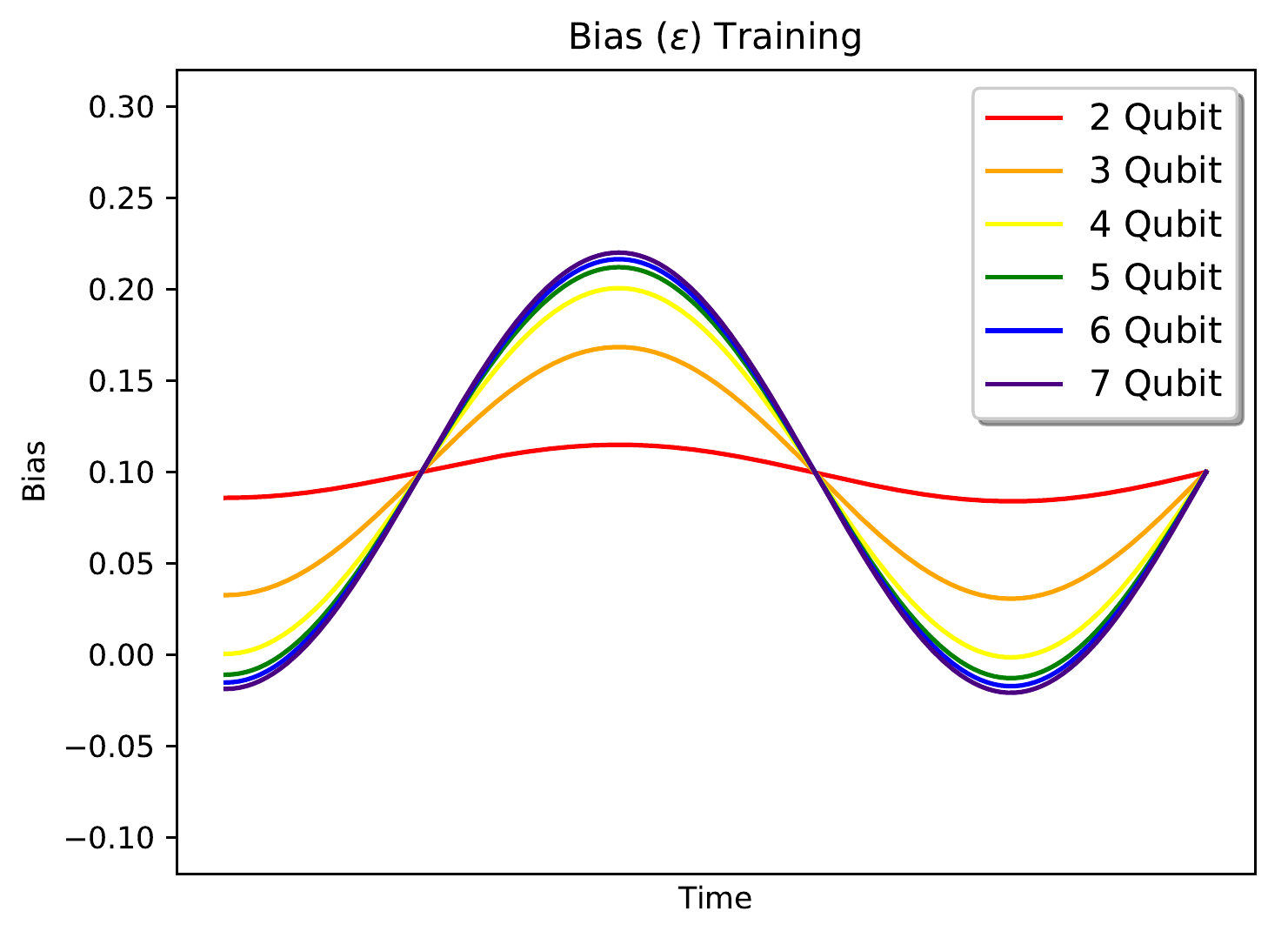}
    \end{center}
        \caption{Trained bias functions for the continuum model of the entanglement witness, for systems of increasing numbers of qubits $N$. Note how the graphs in Figure \ref{Eps_Chunk} are close approximations of the shapes and values of the function for each number of qubits.}
        \label{Eps_Cont}
\end{figure}

Figure \ref{Eps_Chunk} shows the bias as a function of time in the 4 chunk model, for two- through seven-qubit systems, compared with a similar 8 chunk model, where the time chunks or intervals were halved. In each case we see that as the number of qubits increases a similar curve takes shape as the bias function seems to reach asymptotic values. Figure \ref{Eps_Cont} is the continuum case \cite{continuous}, for the same two- through seven-qubit systems. Each system was trained using the techniques outlined in \cite{multiqubit,continuous} with the imposed symmetry constraints as discussed in the previous section. We observe that the 4 and 8 chunk cases show strong qualitative resemblance to the shape of the parameter function in the continuum case. Quantitatively, the exact values of the discretized functions and the continuum model do not match, but the disagreement is small and the overshooting can be attributed to fitting error. In previous work we have shown \cite{2 qubit noise,robust} both that the calculation is relatively insensitive to the exact values of the time dependent functions and that as the system size $N$ is increased that robustness increases; thus the disagreement is probably irrelevant and in any case becomes more irrelevant with increasing $N$. Total RMS error for each of these simulations is shown in Table \ref{48contRMS}.

\begin{table}[h]

\begin{tabular}{c| c c c}
\hline
          &        & RMS error &                 \\
\hline
Qubits    &   4 Chunk & 8 Chunk & Continuous   \\
\hline
2   & $6.0\times 10^{-4}$ & $6.4\times 10^{-4}$ & $5.2\times 10^{-4}$   \\
3   & $6.0\times 10^{-4}$ & $8.8\times 10^{-4}$ & $1.2\times 10^{-3}$   \\
4   & $4.2\times 10^{-4}$ & $1.2\times 10^{-3}$ & $4.3\times 10^{-4}$   \\
5   & $3.6\times 10^{-4}$ & $1.0\times 10^{-3}$ & $4.1\times 10^{-4}$   \\
6   & $1.6\times 10^{-3}$ & $1.2\times 10^{-3}$ & $6.4\times 10^{-4}$   \\
7   & $1.8\times 10^{-3}$ & $1.4\times 10^{-3}$ & $6.6\times 10^{-4}$   \\
\hline
\end{tabular}
\caption{Total RMS error for the training set in the 4, 8, and continuum models of the entanglement witness for system sizes ranging from two- to seven-qubits. Training followed the methods of \cite{multiqubit,continuous} with the additional condition that all parameters are fully symmetric. The continuum model shows the best accuracy, but the discretized versions also trained well and are viable approximations of the continuum model (not realizable in the current hardware.)}
\label{48contRMS}
\end{table}

\section{Conclusions}
\noindent

As an example of using machine learning techniques to train quantum systems to do computations for which no algorithm is known, we have trained a system of qubits to return a witness to its initial pairwise entanglement, by manipulating parameters in a time-dependent Hamiltonian. This procedure is reminiscent of a physical setup like the quantum annealing processors \cite{QA}, which have a time-dependent Hamiltonian (though the parameter flexibility is still severely limited.) But the approach outlined in this paper is a kind of bridge between the annealing and gate approaches to quantum computing: with systematic Hamiltonian design, QA computers could be used as programmable machines as well \cite{LQA}.  The entanglement witness was well approximated by a series of implementable gates. A thorough statistical analysis was done, and a good confidence interval of about 0.0015 is reached after 15,000 shots. The discretized parameter setup models the entanglement witness accurately with respect to two kinds of scaling: increasing the number of qubits and increasing the number of time chunks in the piecewise-constant parameter functions. Agreement was excellent, and the calculation generalizes well and easily as the number of qubits $N$ increases to seven.

Two of the parameter functions so learned seem already to have reached an asymptote, which would mean that the witness could probably be implemented with only small error for much larger values of $N$, or, at minimum, could be trained online with little effort \cite{reinforce}. The qubit-qubit coupling could not definitively be said to have reached its asymptote, but it is at least plausible that a large fraction of the training necessary has already been accomplished, and, again, has reduced the amount of further training necessary. We are currently working on doing exactly that \cite{Thompson2} using automatic differentiation \cite{autoderivatives}.

Physical implementation still poses some problems. There are major limitations in both connectivity and decoherence with the target hardware. For viable hardware implementation, the main consideration is the computational fidelity. Fidelity is lost to both time and inefficient computations. On the available IBM hardware, coherence times are approximately 60 $\mu$s for both depolarization and spin dephasing \cite{Linke}. The time required to apply a single-qubit gate is about 0.130 $\mu$s and two-qubit gates are between 0.250 $\mu$s and 0.450 $\mu$s. Any state preparation and quantum circuit operations must be completed within the 60 $\mu$s interval. Our implementation of the chunked pairwise entanglement witness uses 28 single-qubit and 8 two-qubit gates, which yields a smaller than 8 $\mu$s total time (plus up to 2 $\mu$s to prepare a state); despite this, reproducibility on IBM hardware was not good \cite{AIAA}. Gate fidelity also affects computations. Single qubit readouts are accurate ~96\% of the time, and single- and two-qubits maintain fidelity at a rate of 99.7\% and 96.5\%, respectively \cite{Linke}.  Available hardware and circuit implementation techniques will of course improve. Developers are working on two important avenues to combat decoherence: higher fidelity physical implementation of quantum gates \cite{higherfidelity,dynamicaldecoupling}, and the reduction of the so-called $T$-depth for circuits \cite{GateSynthesis,ancilla}. The value of fidelity increases are obvious, and reducing the physical time required to perform the operations of a circuit will improve computational accuracy. It should be noted that as the coherence times of the hardware improve, our training paradigm increases in value as we can give better models with finer discretizations of the continuum training for our entanglement witness, and, of course, other desired calculations. Moreover, machine learning solutions may also have robustness advantages \cite{2 qubit noise,robust}.

Optimization of the discretization of universal circuits for operators involving very small numbers of qubits at a time is a major advance towards universal quantum computation. But it is not the whole answer. For one thing, many times we do not know the unitary operator that will perform the computation, since we do not have an algorithm. For another, we still do not have optimal ways of reducing an $N$ qubit unitary to building blocks involving only one or two qubits. Machine learning holds a great deal of promise for both tasks. Our work here seems to show that with bootstrapping we can fairly easily extend small simulational results to larger systems. And, even when a unitary is known that performs  the desired calculation, a clever neural network approach may find one with more efficiency or better speedup \cite{GA paper}.

%

\begin{acknowledgements}
We thank Patrick Coles (LANL) and William Ingle (WSU), for helpful discussions, and Henry Elliott (WSU) for the comparative Qiskit \cite{IBMQE} calculations and hardware implementation.
\end{acknowledgements}


\begin{thebibliography}{}
%
%
\bibitem{QA} K. Karimi, N.G. Dickson, F. Hamze, M.H.S. Amin, M. Drew-Brook, F.A. Chudak, P.I. Bunyk, W.G. Macready, and G. Rose, {\it Investigating the performance of an adiabatic quantum optimization processor}, Quantum Inf. Process. {\bf 11}, pp. 77-88 (2012).

\bibitem{shor} P.W. Shor, {\it Algorithms for quantum computation: discrete logarithms and factoring}, Proceedings of the 35th Annual Symposium on Foundations of Computer Science (IEEE) (1994).

\bibitem{grover} L.K. Grover, {\it A fast quantum mechanical algorithm for data base search}, Proceedings of the 28th Annual ACM Symposium on the Theory of Computing 212, (1996).

\bibitem{childs} A.M. Childs, R. Cleve,  E. Deotto,  E. Farhi,  S. Gutmann, and D.A. Spielman, {\it Exponential algorithmic speedup by quantum walk}, Proceedings of the 35th Symposium on Theory of Computing, 59–68 (2003).

\bibitem{bravyi} S. Bravyi, D. Gosset, and R. Konig, {\it Quantum advantage with shallow circuits}, arXiv:1704.00690v1 (2017).

\bibitem{ronnow} T.F. Ronnow, Z. Wang, J. Job, S. Boixo, S.V. Isakov, D. Wecker, J.M. Martinis, D.A. Lidar, and M. Troyer, {\it Defining and detencting quantum speedup}, Science {\bf 345}, 420 (2014).

\bibitem{infsci} E.C. Behrman, L.R. Nash, J.E. Steck, V. Chandrashekar, and S.R. Skinner,  {\it Simulations of quantum neural networks},  Inf. Sci. {\bf 128}, 257 (2000).

\bibitem{previous} E.C. Behrman, V Chandrashekar, Z. Wang, C.K. Belur, J.E. Steck, and S.R. Skinner, {\it A quantum neural network computes entanglement}, arXiv: quant-ph/0202131 (2002).

\bibitem{2008} E.C. Behrman, J.E. Steck, P. Kumar, and K.A. Walsh, {\it Quantum algorithm design using dynamic learning}, Quantum Inf. Comput. {\bf 8} pp. 12-29 (2008).

\bibitem{multiqubit} E. C. Behrman and J. E. Steck, {\it Multiqubit entanglement of a general input state}. Quantum Inf. Comput. {\bf 13},  pp. 36-53 (2013).

\bibitem{bootstrap} B. Efron and R.J. Tibshirani, {\it An introduction to the bootstrap} Boca Raton, FL: Chapman and Hall/CRC (1994).

\bibitem{2 qubit noise} E.C. Behrman, N.H. Nguyen, J.E. Steck, M. McCann, {\it Quantum neural computation of entanglement is robust to noise and decoherence}, in Quantum Inspired Computational Intelligence: Research and Applications, S. Bhattacharyya, ed. (Morgan Kauffman, Elsevier) pp.3-33 (2016).

\bibitem{robust} N.H. Nguyen, E.C. Behrman, and J.E. Steck, {\it Quantum learning with noise and decoherence: a robust quantum neural network},  in {\it Quantum Machine Learning}, Cambridge (to appear); arXiv: 1612.07593 (2019).

\bibitem{related} M. Swaddle, L. Noakes, L. Salter, H. Smallbone, and J. Wang, {\it Generating 3 qubit quantum circuits with neural networks}, Phys. Lett. A {\bf 381}, 3391 (2017).

\bibitem{LQA} E.C. Behrman, J.E. Steck, and M.A. Moustafa, {\it Learning quantum annealing}, Quantum Inf. Comput. {\bf 17}, 0469-0487 (2017).

\bibitem{gurvitz} L. Gurvitz, {\it Classical deterministic complexity of Edmonds problem and quantum entanglement}, in {\it Proc. 35th ACM Symp. on the Theory of Comput.}, pp. 10-19 (ACM Press, New York, 2003).

\bibitem{werbos} P. Werbos, in {\it Handbook of Intelligent Control}, Van Nostrand Reinhold, pp. 79-80 and 339-344 (1992); Yann le Cun, {\it A theoretical framework for back-propagation} in {\it Proc. 1998 Connectionist Models Summer School,} D. Touretzky, G. Hinton, and T. Sejnowski, eds., Morgan Kaufmann, (San Mateo), pp. 21-28 (1988).

\bibitem{neilsen and chuang} M. A. Nielsen and I. L. Chuang, {\it Quantum Computation and Quantum Information}, Cambridge University Press (Cambridge, England) (2001).

\bibitem{lloyd} S. Lloyd, {\it Universal quantum simulators}. Science {\bf 273}, 1073 (1996).

\bibitem{larose} R. LaRose, {\it Overview and comparison of gate level quantum software plaltforms}, arXiv: 1807.02500 (2018).

\bibitem{Qsharp} Microsoft Quantum Development Kit. \url{https://docs.microsoft.com/en-us/quantum/?view=qsharp-preview}, (2018). accessed September 2018

\bibitem{IBMQE} The IBM Quantum Experience. \url{https://quantumexperience.ng.bluemix.net/qx}, (2018).  accessed October 2018

\bibitem{qcompiler} Y.G. Chen and J.B. Wang, {\it QCompiler: quantum compilation with CSF method}, arXiv:quant-ph/1208.0194v2 (2012).

\bibitem{GateSynthesis} V. Kliuchnikov, {\it Synthesis of unitaries with Clifford+ T circuits}, arXiv:1306.3200 (2013);  V. Kliuchnikov, D. Maslow, M. Mosca, {\it Asmptotically optimal approximation of single qubit unitaries by Clifford and T circuits using a constant number of ancillary qubits}. Phys. Rev. Lett. {\bf 110}, 190502 (2013).

\bibitem{pcoles} L. Cincio, Y. Subaşı, A.T. Sornborger, and P.J. Coles, {\it Learning the quantum algorithm for state overlap}, New J. Phys. 20, 113022 (2018).

\bibitem{continuous} E.C. Behrman, R.E.F. Bonde, J.E. Steck, and J.F. Behrman, {\it On the correction of anomalous phase oscillation in entanglement witnesses using quantum neural networks}, IEEE Trans. on Neural Networks and Learning Systems {\bf 25}, pp 1696-1703 (2014).

\bibitem{AIAA} J.E. Steck, E.C. Behrman, and N.L. Thompson, {\it Machine learning applied to programming quantum computers}, AIAA Scitech 2019 Forum, (January 2019). DOI: 10.2514/6.2019-0956

\bibitem{github} Available at https://github.com/williamingle/WichitaStateQNN

\bibitem{preskill} J. Preskill, {\it Quantum computing and the entanglement frontier}, arXiv:1203.5813v3 (2013).

\bibitem{reinforce} M.Y. Niu, S. Boixo, V. Smelyanskiy, and H. Neven,  {\it Universal quantum control through deep reinforcement learning}, arXiv:1803.01857v2 (2018).

\bibitem{Thompson2} N.L. Thompson, N.H. Nguyen, W. Ingle, E.C. Behrman, and J.E. Steck, {\it Optimized quantum machine learning}, in progress (2019).

\bibitem{autoderivatives} M. Schuld, V. Bergholm, C. Gogolin, J. Izaac, and N. Killoran, {\it  Evaluating analytic gradients on quantum hardware,} arXiv:1811.11184 (2018).

\bibitem{Linke} N. M. Linke, D. Maslov, M. Rotteler, S. Debnath, C. Figgatt, K. A. Landsman, K. Wright, and C. R. Monroe, {\it Experimental comparison of two quantum computing architectures,} arXiv:1702.01852 (2017).

\bibitem{higherfidelity} X. Wang, E. Barnes, S.D. Sarma, {\it Improving the gate fidelity of capacitively coupled spin qubits}, NPJ Quantum Information {\bf 1}, 15003, 10.1038/npjqi.2015.3 (2015).

\bibitem{dynamicaldecoupling} B. Pokharel, N. Anand, B. Fortman, D.A. Lidar,{\it Demonstration of fidelity improvement using dynamical decoupling with superconducting qubits}, Phys. Rev. Lett., {\bf 121}, 220502, DOI: 10.1103/PhysRevLett.121.220502 (2016).

\bibitem{ancilla} N.J. Ross and P. Selinger, {\it Optimal ancilla-free Clifford+T approximation of z-rotations}, arXiv: quant-ph/1403.2975 (2014).

\bibitem{GA paper} M.J. Rethinam, A.K. Javali, A.E. Hart, E.C. Behrman, and J.E. Steck,  {\it A genetic algorithm for finding pulse sequences for nmr quantum computing,} Paritantra – Journal of Systems Science and Engineering {\bf 20}, 32-42. arXiv:quant-ph/0404170 (2011).

\end{thebibliography}


\end{document}